\documentclass[aps,pra,preprint,showkeys,superscriptaddress]{revtex4-2}
\usepackage{amsmath}
\usepackage{graphicx}
\begin{document}

%\preprint{}

\title{Stokes parameters alone cannot completely characterize the polarization of plane light waves}


\author{Chun-Fang Li}
\email[Corresponding author. E-mail address: ]{cfli@shu.edu.cn}
%\thanks{}
\affiliation{Department of Physics, Shanghai University, 99 Shangda Road, 200444 Shanghai, China}

\author{Zhi-Juan Hu}
%\email[]{huzhijuan@shnu.edu.cn}
%\thanks{}
\affiliation{Department of Physics, Shanghai Normal University, 100 Guilin Road, 200233 Shanghai, China}


%\date{\today}

\begin{abstract}

It was generally assumed that the Stokes parameters are complete characterization for the state of polarization of a plane light wave so that their counterparts in quantum optics, called the Stokes operators, represent the polarization of photons. Here we show, through analyzing the properties of polarized plane waves in an optically active medium, that the Stokes parameters are not able to completely characterize the state of polarization of a plane wave.
The key point is that only when a plane wave is expanded in terms of the orthogonal base modes, which are physically meaningful, can the two expansion coefficients make up the Jones vector. 
Taking this into consideration, we demonstrate that the Stokes parameters of any elliptically polarized wave in an isotropic chiral medium, determined solely by its Jones vector, are transmitted unchanged. They are not able to reflect the rotation of its polarization ellipse along with the propagation.
The relationship of the Stokes parameters with the polarization of light needs further investigation.

\end{abstract}



\maketitle


\newpage

\section{Introduction}

Polarization is one of the most fundamental phenomena of light. It arises from the vector nature of the electromagnetic wave that is governed by Maxwell's equations. The significant point is that apart from satisfying the coupled equations, the electric field $\mathbf E$ and magnetic field $\mathbf H$ are also constrained by the transversality conditions,
\[\nabla \cdot \mathbf{E}(\mathbf{x},t)=0, \quad \nabla \cdot \mathbf{H}(\mathbf{x},t) =0.\]
So, there have developed two main mathematical descriptions of the polarization \cite{Hecht}. 
One makes use of the Stokes parameters that were introduced in 1852 \cite{Stok}. They are connected with the polarization matrix \cite{Wolf, Mart-MP} in classical optics or with the density matrix \cite{Falk-M,Fano} in quantum mechanics. 
The other is the so-called Jones vector introduced in 1941 \cite{Jone}. They are, nevertheless, not equivalent. The former is applicable to partially polarized waves. The latter is only applicable to completely polarized waves. However, for a monochromatic plane wave, they have a definite and simple relationship \cite{Dama, Gold-HB, Merz}.
It was generally assumed that the Stokes parameters, as physically observable quantities \cite{Fano49, Coll70, Gold}, provide a full characterization of the state of polarization, at least for plane waves or beamlike waves \cite{McMa, Jame-KMW, Elli-DPW}. 
Here we will show that this is actually not the case. 
We will do so by demonstrating that the Stokes parameters cannot characterize the rotation of the polarization ellipse of the plane wave in an isotropic chiral medium.

A chiral medium is an optically active medium \cite{Barr}. It has the ability to rotate the polarization ellipse of an elliptically polarized wave that passes through it \cite{Lind-STV, Coll}. 
The conventional description of the optical activity is the circular birefringence, proposed in 1825 by Fresnel \cite{Hecht}. By circular birefringence it is meant that the right-handed circularly polarized and left-handed circularly polarized waves in a chiral medium propagate at different velocities.
Recently we showed \cite{Hu-L} through a logical analysis that such a phenomenological description of the optical activity is incorrect. Circular birefringence does not exist in chiral media. Meanwhile we found that the rotation of the polarization ellipse of the elliptically polarized wave in chiral media lies with the rotation of the polarization basis, without involving the change of the Jones vector. In particular, we demonstrated that the rotation of the right-handed circularly polarized and left-handed circularly polarized waves gives rise to phases of equal magnitude but opposite sign as if they propagated at different phase velocities with their polarization states transmitted unchanged \cite{Ditc}.
The aim of this paper is to show that the Stokes parameters determined by the Jones vector cannot reflect the rotation of the polarization ellipse. The reported result raises the problem of how to interpret the physical meaning of the Stokes parameters.

\section{Conventional theory of the polarization of light}

Before discussing the peculiarity of the Stokes parameters of a plane wave in chiral media, it is beneficial to overview how the Stokes parameters of a plane light wave in ordinary achiral media are defined. 
Because the intensity of a monochromatic plane wave has nothing to do with the state of its polarization, we will consider only electric field of normalized amplitude.
Letting a plane wave of frequency $\omega$ and wavenumber $k$ propagate along the $z$-axis, the electric field can be written as
\begin{equation}\label{E-AM}
	\mathbf{E}^{(A)}(z,t)=\mathbf{a}^{(A)} \exp[i(kz-\omega t)],
\end{equation}
where the normalized polarization vector $\mathbf{a}^{(A)}$ takes the form of
\begin{equation}\label{PV}
	\mathbf{a}^{(A)} =\alpha_1 \bar{x} +\alpha_2 \bar{y},
\end{equation}
$\bar x$ and $\bar y$ are the unit vectors along the $x$- and $y$-axes, respectively, and the complex coefficients $\alpha_1$ and $\alpha_2$ satisfy the normalization condition
$|\alpha_1|^2 +|\alpha_2|^2 =1$. 
The Stokes parameters of the wave are defined as follows \cite{Dama, Gold},
\begin{equation}\label{SP}
	s_i =\alpha^\dag \hat{\sigma}_i \alpha, \quad i=1,2,3,
\end{equation}
where 
$
\alpha =\bigg(\begin{array}{c}
     	          \alpha_1 \\ \alpha_2
              \end{array}
        \bigg)
$
is known as the Jones vector of the wave, the superscript $\dag$ denotes conjugate transpose, and
\begin{equation*}
	\hat{\sigma}_1=\bigg(\begin{array}{cc}
		1 &  0 \\
		0 & -1
	\end{array}
	\bigg),                \quad
	\hat{\sigma}_2=\bigg(\begin{array}{cc}
		0 & 1 \\
		1 & 0
	\end{array}
	\bigg),                \quad
	\hat{\sigma}_3=\bigg(\begin{array}{cc}
		0 & -i \\
		i &  0
	\end{array}
	\bigg)
\end{equation*}
are the Pauli matrices. 
The unit vectors $\bar x$ and $\bar y$ in Eq. (\ref{PV}) are commonly referred to as the polarization base vectors.
It is important to point out, as is known in the literature \cite{Falk-M,Gold}, that when the Stokes parameters are defined in terms of the Jones vector via Eq. (\ref{SP}), it is implied that the plane wave (\ref{E-AM}) is a superposition of the following two base modes,
\begin{equation*}
	\begin{split}
		\mathbf{E}^{(A)}_1 (z,t) &=\bar{x} \exp[i(kz-\omega t)], \\
		\mathbf{E}^{(A)}_2 (z,t) &=\bar{y} \exp[i(kz-\omega t)].
	\end{split}
\end{equation*}
That is, the polarization base vectors $\bar x$ and $\bar y$ are polarization vectors of these base modes, which are physically meaningful.
This idea is more understandable from the point of view of quantum mechanics, where one deals with orthonormal base states \cite{Fano49, Coll70, McMa}. For later convenience, we cast Eq. (\ref{PV}) into a compact form \cite{Li},
\begin{equation}\label{PV-varpi}
	\mathbf{a}^{(A)} =\varpi^{(A)} \alpha,
\end{equation}
where the convention of matrix multiplication is used and the row matrix
$\varpi^{(A)}=(\begin{array}{lr}
	               \bar{x} & \bar{y}
               \end{array} )
$
represents the polarization basis $\bar x$ and $\bar y$. 

Any two orthonormal polarization vectors may be chosen as the polarization basis. Apart from the linear-polarization basis, another common basis is the pair of circularly polarized states \cite{Gold-HB, McMa, Bhan, Merz}. In terms of the circular-polarization basis, which is related to the linear-polarization basis via
\begin{equation}\label{RT} %representation transformation
	\mathbf{R} =\frac{1}{\sqrt 2} (\bar{x} +i\bar{y}), \quad
	\mathbf{L} =\frac{1}{\sqrt 2} (\bar{x} -i\bar{y}),
\end{equation}
the same polarization vector (\ref{PV}) can be expanded as
\begin{equation}\label{PV-C}
	\mathbf{a}^{(A)} = \alpha_R \mathbf{R} + \alpha_L \mathbf{L},
\end{equation}
where
\begin{equation}\label{UT-C}
	\alpha_R =(\alpha_1 -i\alpha_2)/\sqrt{2}, \quad
	\alpha_L =(\alpha_1 +i\alpha_2)/\sqrt{2}.
\end{equation}
Equations (\ref{RT}) describe a transformation of representation for the state of polarization in the language of quantum mechanics. In this case, the same Stokes parameters (\ref{SP}) can be rewritten as
\begin{equation}\label{SP2}
s_i = \alpha^{c\dag} \hat{\sigma}^c_i \alpha^c ,
\end{equation}
where 
$
\alpha^c =\bigg(\begin{array}{c}
	          \alpha_R \\ \alpha_L
                \end{array}
          \bigg)  
$ is the Jones vector in the representation of circular-polarization basis and
\begin{equation*}
	\hat{\sigma}^c_1=\bigg(\begin{array}{cc}
		0 & 1 \\
		1 & 0
	\end{array}
	\bigg),                \quad
	\hat{\sigma}^c_2=\bigg(\begin{array}{cc}
		0 & -i \\
		i &  0
	\end{array}
	\bigg),                \quad
	\hat{\sigma}^c_3=\bigg(\begin{array}{cc}
		1 &  0 \\
		0 & -1
	\end{array}
	\bigg).            
\end{equation*}
This is understandable. According to Eqs. (\ref{UT-C}), $\alpha^c$ is related to $\alpha$ via
$\alpha^c =M \alpha$, where
$
M=\frac{1}{\sqrt 2} \bigg(\begin{array}{cc}
	1 & -i \\
	1 &  i
\end{array}
\bigg)
$ is a unitary matrix, so that
\[ \alpha=M^\dag \alpha^c. \]
Substituting it into Eq. (\ref{SP}), one will arrive at Eq. (\ref{SP2}) in which $\hat{\sigma}^c_i$ is related to $\hat{\sigma}_i$ via
$\hat{\sigma}^c_i =M \hat{\sigma}_i M^\dag$. After all, the Stokes parameters as physically observable quantities do not depend on the choice of representation.
Of course, expression (\ref{SP2}) for the Stokes parameters implies that the polarization base vectors $\mathbf{R}$ and $\mathbf{L}$ are the polarization vectors of physically meaningful states. In other words, expression (\ref{PV-C}) for the polarization vector implies that the plane wave (\ref{E-AM}) can be expanded in terms of the following circularly polarized base states,
\begin{equation*}
	\begin{split}
		\mathbf{E}^{(A)}_R (z,t) &=\mathbf{R} \exp[i(kz-\omega t)], \\
		\mathbf{E}^{(A)}_L (z,t) &=\mathbf{L} \exp[i(kz-\omega t)].
	\end{split}
\end{equation*}

We note that because the polarization base vectors $\bar x$ and $\bar y$ in expression (\ref{PV}) are fixed, the Jones vector $\alpha$ is equivalent to the polarization vector $\mathbf{a}^{(A)}$ in describing the state of polarization. For this reason, Eq. (\ref{PV}) or (\ref{PV-varpi}) is usually simplified as \cite{Hecht, Mart-MP, Gold}
\begin{equation}\label{PV=JV}
	\mathbf{a}^{(A)} =\alpha.
\end{equation}
The change in the polarization vector amounts to the change in the Jones vector,
\begin{equation*}
	\mathbf{a}^{(A)} \rightarrow \mathbf{a}'^{(A)}=\alpha'_1 \bar{x} +\alpha'_2 \bar{y}
	                 =\bigg(\begin{array}{c}
	                 	        \alpha'_1 \\ \alpha'_2
	                 	    \end{array}     
	                  \bigg) \equiv \alpha'.
\end{equation*}
When the energy is conserved, the change in the Jones vector is described by a unitary transformation \cite{Gold-HB},
\[ \alpha'=U \alpha, \]
where $U$ is a $2 \times 2$ unitary matrix, which can be expressed as a matrix of rotation of an angle $\varphi$ about an axis represented by a real-valued unit vector $\mathbf n$,
\begin{equation}\label{RoSU2}
	U(\varphi,\mathbf{n})= \exp(-i \mathbf{n} \cdot \boldsymbol{\sigma} \varphi/2).
\end{equation}
It is probably on this basis that the Stokes parameters determined by the Jones vector are believed to provide a complete characterization for the state of polarization represented by the polarization vector.
But unfortunately, the plane wave in chiral media no longer has a fixed polarization basis as we will show below.

\section{Stokes parameters of plane waves in a chiral medium}

Now we examine the properties of the Stokes parameters of plane waves in a chiral medium. The optical property of a chiral medium is conveyed by its constitutive relations. The constitutive relations of an isotropic and transparent chiral medium can be written as follows \cite{Silv,Lekn},
\begin{equation}\label{CR}
	\mathbf{D} =\varepsilon \mathbf{E}-g \partial{\mathbf{H}}/\partial{t}, \quad
	\mathbf{B} =\mu \mathbf{H}+g \partial{\mathbf{E}}/\partial{t},
\end{equation}
where $\mathbf D$ and $\mathbf B$ are, as usual, the vectors of electric displacement and magnetic induction, respectively, $\varepsilon$ is the permittivity, $\mu$ is the permeability, and the pseudo-scalar constant $g$ is the gyrotropic parameter.
The same as above, we assume that the plane wave of angular frequency $\omega$ in the chiral medium propagates along the $z$-axis. Constitutive relations (\ref{CR}) allow one to find, from Maxwell's equations, the following two orthonormal circularly polarized waves \cite{Hu-L},
\begin{equation}\label{Er-El}
	\begin{split}
		\mathbf{E}_R (z,t) &=\mathbf{R} \exp(-i\tau z) \exp[i(kz-\omega t)], \\
		\mathbf{E}_L (z,t) &=\mathbf{L} \exp( i\tau z) \exp[i(kz-\omega t)],
	\end{split}
\end{equation}
where $\mathbf R$ and $\mathbf L$ are given by Eqs. (\ref{RT}), $k=(\varepsilon \mu)^{1/2} \omega$, and $\tau=-g \omega^2$. These two circularly polarized waves can be taken as base states to expand any elliptically polarized wave \cite{Lind-STV},
\begin{equation}\label{E-EP}
	\mathbf{E}(z,t)=\alpha_R \mathbf{E}_R (z,t) +\alpha_L \mathbf{E}_L (z,t),
\end{equation}
where the expansion coefficients $\alpha_R$ and $\alpha_L$, which are constant, satisfy the normalization condition
$|\alpha_R|^2 +|\alpha_L|^2 =1$.
Now that $\mathbf{E}_R$ and $\mathbf{E}_L$ are two orthonormal base states, $\alpha_R$ and $\alpha_L$ make up the Jones vector of the wave (\ref{E-EP}),
$
\alpha^c =\bigg(\begin{array}{c}
	                \alpha_R \\ \alpha_L
                \end{array}
          \bigg)  
$,
in accordance with Refs. \cite{Falk-M, Gold, McMa, Fano49, Coll70}.
According to Eq. (\ref{SP2}), the Stokes parameters determined by this Jones vector are constants, independent of the propagation distance $z$. 
However, the polarization ellipse of the wave (\ref{E-EP}) is rotated along with the propagation. 
We are thus convinced that the Stokes parameters are not able to characterize the rotation of the polarization state of the plane wave in the chiral medium.

Let us look at this result in the representation of linear-polarization basis. As is well known, Eq. (\ref{E-EP}) can give rise to linearly polarized waves \cite{Hecht,Gold}. 
In particular, it can give rise to two orthonormal linearly polarized waves. 
Specifically, if $\alpha_R=\alpha_L=\frac{1}{\sqrt 2}$, Eq. (\ref{E-EP}) represents a linearly polarized wave, denoted by 
\begin{equation}\label{E1}
	\mathbf{E}_1 =\frac{\mathbf{E}_R +\mathbf{E}_L}{\sqrt 2}
	             =\mathbf{a}_1 (z) \exp[i(kz-\omega t)],
\end{equation}
where 
\begin{equation}\label{a1}
\mathbf{a}_1 (z) =\bar{x} \cos \tau z +\bar{y} \sin \tau z.
\end{equation}
In addition, if $\alpha_R=-\alpha_L=-\frac{i}{\sqrt 2}$, it also represents a linearly polarized wave, denoted by
\begin{equation}\label{E2}
	\mathbf{E}_2 =\frac{\mathbf{E}_R -\mathbf{E}_L}{i \sqrt 2}
	             =\mathbf{a}_2 (z) \exp[i(kz-\omega t)],
\end{equation}
where
\begin{equation}\label{a2}
\mathbf{a}_2 (z) =\bar{y}\cos\tau z -\bar{x}\sin\tau z.
\end{equation}
Being rotated along with the propagation, the polarization vectors of these two linearly polarized waves, (\ref{a1}) and (\ref{a2}), are both $z$-dependent. They are, however, orthogonal to each other at the same propagation distance $z$,
\[ \mathbf{a}_1 (z) \cdot \mathbf{a}_2 (z) =0. \]
So, these two linearly polarized waves can also be taken as base states to expand any elliptically polarized wave,
\begin{equation}\label{E-EP3}
	\mathbf{E} (z,t)=\alpha_1 \mathbf{E}_1 +\alpha_2 \mathbf{E}_2 ,
\end{equation}
where the constant expansion coefficients $\alpha_1$ and $\alpha_2$ make up the Jones vector
$
\alpha =\bigg(\begin{array}{c}
	\alpha_1 \\ \alpha_2
\end{array}
\bigg)
$, 
which satisfies the normalization condition
$\alpha^\dag \alpha=1$.
According to Eq. (\ref{SP}), the Stokes parameters determined by this Jones vector are also independent of the propagation distance. 
In fact, if expression (\ref{E-EP3}) stands for the same wave as expression (\ref{E-EP}), we must have
\begin{equation*}
	\alpha_R=(\alpha_1 -i\alpha_2)/\sqrt{2}, \quad 
	\alpha_L=(\alpha_1 +i\alpha_2)/\sqrt{2},
\end{equation*}
by virtue of Eqs. (\ref{E1}) and (\ref{E2}).
They are the same as Eqs. (\ref{UT-C}).
In a word, the Stokes parameters are not able to characterize the rotation of the polarization state of a plane wave in the chiral medium.

\section{Explanations and discussions}

The above result can be explained with the newly-advanced description \cite{Hu-L} of the optical activity.
As is well known, the planes of polarization of the linearly polarized waves (\ref{E1}) and (\ref{E2}) are rotated with propagation. To illustrate this more clearly, we rewrite their polarization vectors (\ref{a1}) and (\ref{a2}) as follows,
\begin{equation}\label{a12}
	\mathbf{a}_1 (z) =\exp[-i(\bar{z} \cdot \mathbf{\Sigma}) \tau z] \bar{x}, \quad
	\mathbf{a}_2 (z) =\exp[-i(\bar{z} \cdot \mathbf{\Sigma}) \tau z] \bar{y},
\end{equation}
where $(\Sigma_k)_{ij} =-i \epsilon_{ijk}$ with $\epsilon_{ijk}$ the Levi-Civit\'{a} pseudotensor and $\bar z$ denotes the unit vector along the $z$-axis. To give Eqs. (\ref{a12}), we have made use of the formula \cite{Norm}
\begin{equation*}
	\exp[-i(\mathbf{a} \cdot \mathbf{\Sigma}) \phi] \mathbf{b}
	=\mathbf{b} \cos \phi -i (\mathbf{a} \cdot \mathbf{\Sigma}) \mathbf{b} \sin \phi 
	+\mathbf{a} (\mathbf{a} \cdot \mathbf{b}) (1-\cos \phi)
\end{equation*}
and the equlity \cite{Schi}
\begin{equation*}
	(\mathbf{a} \cdot \mathbf{\Sigma}) \mathbf{b} =i \mathbf{a} \times \mathbf{b},
\end{equation*}
where $\mathbf a$ and $\mathbf b$ are any two vectors. 
It is seen that the polarization vectors $\mathbf{a}_1$ and $\mathbf{a}_2$ result from the same rotation of $\bar x$ and $\bar y$, respectively. This is why they are orthogonal to each other at the same propagation distance $z$. 
Substituting Eqs. (\ref{E1}) and (\ref{E2}) into Eq. (\ref{E-EP3}), we get
\begin{equation}\label{E-LPB}
	\mathbf{E} (z,t) =[\alpha_1 \mathbf{a}_1 (z) +\alpha_2 \mathbf{a}_2 (z)] \exp [i(kz-\omega t)].
\end{equation}
Now that $\mathbf{a}_1$ and $\mathbf{a}_2$ are polarization vectors of physically meaningful states (\ref{E1}) and (\ref{E2}), respectively, and are orthogonal to each other, they act in this expression as the linear-polarization basis. So there is no doubt that the coefficients $\alpha_1$ and $\alpha_2$ make up the Jones vector $\alpha$. 
What is noteworthy here is, as just mentioned, that these polarization base vectors are simultaneously rotated with propagation, in sharp contrast with those in expression (\ref{PV}) for the polarization vector in the ordinary achiral medium.
This shows that the optical activity described by the polarization vector
\begin{equation*}
	\mathbf{a}(z)=\alpha_1 \mathbf{a}_1 (z) +\alpha_2 \mathbf{a}_2 (z)
\end{equation*}
comes from the rotation of the polarization basis other than from the change of the Jones vector. 

To express this peculiarity explicitly, we cast the above equation for the polarization vector into a compact form,
\begin{equation}\label{QUT}
	\mathbf{a}(z)=\varpi(z) \alpha,
\end{equation}
where the row matrix
$\varpi(z)=(\begin{array}{lr}
	            \mathbf{a}_1 & \mathbf{a}_2
            \end{array}
           )
$, 
which consists of the unit vectors $\mathbf{a}_1 (z)$ and $\mathbf{a}_2 (z)$, represents the linear-polarization basis. It indicates that the polarization vector $\mathbf{a}(z)$ at any propagation distance $z$ has its own polarization basis $\varpi(z)$ with the Jones vector $\alpha$ being the same. 
Obviously, we have
\begin{equation}\label{varpi}
	\varpi(z) =\exp[-i(\bar{z} \cdot \mathbf{\Sigma}) \tau z] \varpi(0)
\end{equation}
in accordance with Eqs. (\ref{a12}), where
$\varpi(0)=(\begin{array}{lr}
	            \bar{x} & \bar{y}
            \end{array}
           )
$
is the polarization basis at $z=0$.
Due to the rotation of the polarization basis with propagation, the Jones vector $\alpha$ here is no longer equivalent to the polarization vector $\mathbf{a}(z)$. That is to say, Eq. (\ref{QUT}) cannot be simplified as 
\[ \mathbf{a} (z) =\alpha \]
in the way that Eq. (\ref{PV}) is simplified as Eq. (\ref{PV=JV}). This is why the Stokes parameters determined by the Jones vector via Eq. (\ref{SP}) are not able to characterize the rotation of the state of polarization.
It is remarked that the rotation of the polarization basis given by Eqs. (\ref{a12}) is not to be confused with Eqs. (\ref{RT}), which describe a transformation of the representation. 

By the way, it is pointed out that the state of polarization of light in a particular chiral medium is always rotated in the same way no matter what its Jones vector is. In fact, substituting Eq. (\ref{varpi}) into Eq. (\ref{QUT}), we have
\begin{equation}\label{RoA}
    \mathbf{a}(z) =\exp[-i(\bar{z} \cdot \mathbf{\Sigma}) \tau z] \mathbf{a}(0),
\end{equation}
where
$\mathbf{a}(0) =\varpi(0) \alpha$
is the polarization vector at $z=0$. Eq. (\ref{RoA}) shows that the angle of rotation of the polarization ellipse per unit length is $\tau$ regardless of the Jones vector. 
In the cases of circular polarization, the result of rotation turns out to be $z$-dependent phase factors \cite{Hu-L} as is explicitly expressed in Eqs. (\ref{Er-El}).

It is also noted that substitution of Eqs. (\ref{a1}) and (\ref{a2}) into Eq. (\ref{E-LPB}) gives
\begin{equation}\label{E-EP2}
	\mathbf{E}(z,t)=[\alpha_x (z) \bar{x} +\alpha_y (z) \bar{y}] \exp[i(kz-\omega t)],
\end{equation}
where
\begin{equation*}
	\begin{split}
		\alpha_x (z) & =\alpha_1 \cos \tau z -\alpha_2 \sin \tau z,\\
		\alpha_y (z) & =\alpha_1 \sin \tau z +\alpha_2 \cos \tau z.
	\end{split}
\end{equation*}
If the fixed unit vectors $\bar x$ and $\bar y$ were taken as the polarization basis in the same way as is done in the ordinary achiral medium, the coefficients $\alpha_x$ and $\alpha_y$ would make up a $z$-dependent Jones vector \cite{Dama},
$
\alpha'(z) =\bigg(\begin{array}{c}
	                  \alpha_x \\ \alpha_y
                  \end{array}
            \bigg)
$.
From Eq. (\ref{SP}) it follows that the Stokes parameters defined by this ``Jones vector'', 
$s'_i =\alpha'^\dag \hat{\sigma}_i \alpha'$, are given by
\begin{subequations}
	\begin{align}
		s'_1 =&(\alpha^*_1 \alpha_1 -\alpha^*_2 \alpha_2) \cos2\tau z 
		     -(\alpha^*_1 \alpha_2 +\alpha^*_2 \alpha_1) \sin2\tau z, \label{s'1} \\
		s'_2 =&(\alpha^*_1 \alpha_1 -\alpha^*_2 \alpha_2) \sin2\tau z 
		     +(\alpha^*_1 \alpha_2 +\alpha^*_2 \alpha_1) \cos2\tau z, \label{s'2} \\
		s'_3 =&-i(\alpha^*_1 \alpha_2 -\alpha^*_2 \alpha_1),
	\end{align}
\end{subequations}	
which are usually $z$-dependent. Nevertheless, so defined Stokes parameters do not correctly reflect the rotation of the state of polarization.
This is because in the cases of circular polarization, 
$\alpha =\frac{1}{\sqrt 2}
         \bigg(\begin{array}{c}
         	       1 \\ \pm i
               \end{array}    
         \bigg)$,
the Stokes parameters given by these equations are not $z$-dependent,
\[ s'_1 =0, \quad s'_2 =0, \quad s'_3 =\pm 1. \]
They are not able to convey the rotation of the circularly polarized waves as we just mentioned. The key point is that the state of polarization of a linearly polarized wave in the chiral medium is rotated with propagation. Its polarization vector at any propagation distance $z$ can no longer be the fixed $\bar x$ or $\bar y$.
So, even though the electric field of any elliptically polarized wave in the chiral medium can be written mathematically in the form of Eq. (\ref{E-EP2}), it is physically unreasonable to take $\bar x$ and $\bar y$ as the polarization basis. This is to say that $\alpha'(z)$ cannot be a physically meaningful Jones vector at all.

\section{Conclusions and Remarks}\label{conclusions}

In conclusion, we showed, through analyzing the optical rotation in an isotropic chiral medium, that the Stokes parameters determined by the Jones vector via Eq. (\ref{SP}) are not able to completely characterize the state of polarization of plane light waves.
We demonstrated that the polarization vector of the plane wave in the chiral medium can no longer be equivalent to its Jones vector, in contrast with the polarization vector of the plane wave in the ordinary achiral medium. The key difference is that in the latter case the fixed unit vectors $\bar x$ and $\bar y$ in the laboratory reference frame can always be taken as the polarization basis, whereas in the former case no such polarization basis exists.
However, Eq. (\ref{QUT}) indicates that the Stokes parameters determined by the Jones vector $\alpha$ are indeed physical quantities to describe the state of polarization of the plane wave at any propagation distance $z$ in the chiral medium, though they are the same at different values of $z$. Therefore, in order to completely characterize the rotation of the polarization of this wave along with the propagation, an extra degree of freedom is needed. This is the rotation angle $\tau z$ that characterizes the rotation of the polarization basis (\ref{varpi}) along with the propagation. 
In a word, we discovered from the optical activity that the Stokes parameters are not able to completely characterize the state of polarization of plane waves.

This  result raises a significant question about the nature of the Stokes parameters themselves. As a matter of fact, the statements about the nature of the Stokes parameters in the literature are rather subtle.
On one hand, they are theoretically described as quantities in an abstract three-dimensional Euclidean space \cite{Merz}, which is sometimes called Stokes space \cite{Dama}. On the other hand, they are considered to be experimentally observable \cite{Fano49, Coll70, Gold}. It is hard to conceive that a physical quantity in an abstract space can be observed. 
The fact is that when they are measured experimentally \cite{Hecht, Gold, Berr-GL, Scha-CSBF},
the Stokes parameters defined via Eqs. (\ref{SP}) and (\ref{PV-varpi}) are always treated as quantities in the laboratory reference frame. This is because what the linear-polarization basis $\varpi^{(A)}$ in Eq. (\ref{PV-varpi}) represents is exactly the laboratory reference frame \cite{Fano} when the plane wave is assumed to propagate along the $z$-axis. The unit vector $\mathbf n$ in Eq. (\ref{RoSU2}) is with respect to this reference frame.

But as pointed out by Messiah \cite{Mess} and Mandel and Wolf \cite{Mand-W}, even the linear-polarization basis for plane waves in the achiral medium is undefined up to a rotation about the propagation direction. 
One can choose, within the laboratory reference frame represented by $\varpi^{(A)}$, a new linear-polarization basis, say
$\varpi'^{(A)}=(\begin{array}{lr}
 	                \bar{x}' & \bar{y}'
                \end{array} )
$,
where the unit vectors $\bar{x}'$ and $\bar{y}'$ together with the propagation direction, the $z$-axis, form a new right-handed Cartesian frame.
In such a case, the resultant Stokes parameters of the same plane wave (\ref{E-AM}) will be quantities in the new reference frame rather than in the laboratory reference frame.
Noticing such a relationship, we have reason to believe that the Stokes parameters defined via Eqs. (\ref{SP}) and (\ref{QUT}) are no longer quantities in the laboratory reference frame. Instead, they correspond, at any propagation distance, to a local reference frame represented by the polarization basis $\varpi(z)$. The rotation of the polarization basis implies the rotation of the local reference frame.
This may help understand why the Stokes parameters are not able to completely characterize the state of polarization of plane waves. After all, whether polarization vector (\ref{PV-varpi}) in the achiral medium or polarization vector (\ref{QUT}) in the chiral medium is quantity in the laboratory reference frame.
If that is the case, what is the physical meaning of the Stokes parameters themselves in the local reference frame? How are they related to the state of polarization in the laboratory reference frame? Discussions of these issues are beyond the topic of this paper and will be presented elsewhere.

%\section*{Declaration of competing interest}
%The authors declare that they have no known competing financial interests or personal relationships that could have appeared to influence the work reported in this paper. 

%\section*{Data availability}
%No data was used for the research described in the article.

\section*{Acknowledgments}

This work was supported in part by National Natural Science Foundation of China under Grant No. 11974251.

\end{document}